\documentclass[10pt,letterpaper]{article}
\usepackage[top=0.95in,left=1.25in]{geometry}

\usepackage{changepage}

\usepackage[utf8]{inputenc}

\usepackage{textcomp,marvosym}

\usepackage{fixltx2e}

\usepackage{amsmath,amssymb}

\usepackage{cite}

\usepackage{nameref,hyperref}

\usepackage[right]{lineno}

\usepackage{microtype}
\DisableLigatures[f]{encoding = *, family = * }

\usepackage{rotating}

\raggedright
\usepackage[aboveskip=1pt,labelfont=bf,labelsep=period,justification=raggedright,singlelinecheck=off]{caption}

\makeatletter
\renewcommand{\@biblabel}[1]{\quad#1.}
\makeatother

\date{}

\usepackage{lastpage,fancyhdr,graphicx}
\fancyheadoffset[L]{2.25in}
\fancyfootoffset[L]{2.25in}
\begin{document}
\vspace*{0.35in}

\begin{flushleft}
{\Large
\textbf\newline{Article Title - Stimulus-Dependent Frequency Modulation of Information Transmission in Neural Systems}
}
\newline
\\
Lianchun Yu\textsuperscript{1,2*},
Longfei Wang\textsuperscript{1},
Fei Jia\textsuperscript{3},
Duojie Jia\textsuperscript{4}
\\
\bf{1} Institute of Theoretical Physics, Lanzhou University,
Lanzhou 730000, China
\\
\bf{2} Key Laboratory for Magnetism and Magnetic Materials of the Ministry of Education, Lanzhou University, Lanzhou 730000, China
\\
\bf{3} Cuiying College, Lanzhou University,
Lanzhou 730000, China
\\
\bf{4} Institute of Theoretical Physics, College of Physics and Electronic Engineering, Northwest Normal University, Lanzhou 730070, China

%
%





* E-mail: yulch@lzu.edu.cn
\end{flushleft}
\section*{Abstract}
Neural oscillations are universal phenomena and can be observed at different levels of neural systems, from single neuron to macroscopic brain. The frequency of
those oscillations are related to the brain functions. However, little is know about how the oscillating frequency of neural system affects neural information transmission in them. In this paper, we investigated how the signal processing in single neuron is modulated by subthreshold membrane potential oscillation generated by upstream rhythmic neural activities. We found that the high frequency oscillations facilitate the transferring of strong signals, whereas slow oscillations the weak signals. Though the capacity of information convey for weak signal is low in single neuron, it is greatly enhanced when weak signals are transferred by multiple pathways with different oscillation phases.
We provided a simple phase plane analysis to explain the mechanism for this stimulus-dependent frequency modulation in the leakage integrate-and-fire neuron model. Those results provided a basic understanding of how the brain could modulate its information processing simply through oscillating frequency.

%

\section*{Introduction}
The rhythmic or repetitive neural activity in the central nervous system, often termed as ``neural oscillation'', are observed throughout the central nervous system and at all levels, e.g., spike trains, local field potentials and large-scale oscillations~\cite{Buzsakibook, Levine1999book}. Neural oscillation can be generated in many ways, driven either by
mechanisms localized within individual neurons or by interactions between neurons~\cite{Buzsaki2004, Destexhe1993, Buzsaki2003}. In individual neurons, oscillations appear either as oscillations in
 membrane potential or as rhythmic patterns of action potentials, which can cause oscillatory activation of post-synaptic neurons\cite{citeulike:10600957}. At the level of neural
ensembles, synchronized activity of large numbers of neurons can give rise to macroscopic oscillations, which can be observed in the electroencephalogram (EEG)~\cite{Niedermeyer, Wolfgang1999}.

The possible roles of oscillatory neural activity in cognition are revealed by their task dependent changes of frequency~\cite{Ward2003, basar2000,Wangxiaojing2010PhysiolRev}.  For example, theta waves with a lower frequency range, usually around $6$-$7$ Hz, are sometimes observed when a rat is motionless but
 alert, whereas faster rhythms such as gamma activity have been linked to cognitive processing~\cite{Kramis197558,Vanderwolf2000217, Miltner1999}. What's more, the abnormalities of specific frequency bands are related to brain dysfunctions and brain diseases~\cite{Uhlhaas2010}. For example, alterations in gamma band, often accompanied with the abnormalities of lower frequencies (theta or alpha) was reported in Schizophrenia patients~\cite{Lauren2011}.

However, little is known how the neural oscillation play a role in the neural computation or information processing in the lower level, such as neuronal circuit or even single neuron. Many studies have shown that anatomical connectivity is fundamental to the neural computation, such as working memory~\cite{Mongillo2008, Compte2000}. Short and long term synaptic plasticity may serve as a mechanism to modify the connections on a range of time scale from sub-second to lifelong~\cite{Tetzlaff2013}. This modification could provide the flexibility to the brain to make the animals' behaviors adjust to the environment changes~\cite{Akam2012}. Though true of those facts, there is possibility that rapid and selective modification of signal processing may be achieved by using oscillatory dynamics rather than directly modifying synaptic connections~\cite{Akam2010}.

In this paper, we investigated how the frequency of upstream rhythmic activity modulated the information transferring capacity of downstream neurons
 by inducing subthreshold membrane potential oscillation in them. The subthreshold membrane potential oscillation, with their frequency varying from few Hz to over 40 Hz,  have been shown to control action potential timing, facilitate synchronous activity, and extend the integration window for synaptic input~\cite{Desmaisons1999,Lampl1993}. Though the subthreshold oscillations can also be caused by intrinsic electrical properties of neurons~\cite{Chapman1999}, this presynaptic activity caused oscillation is particular interesting because it reflected how the information processing in single neuron is modulated by the upstream neuron through oscillatory activities. We use the information rate calculated with `direct method' to measure the information
 convey capacity of the spike trains generated by a leakage integrate-and-fire(LIF) neuron in response to synaptic pulse inputs. We found that for a single neuron, its information rate
depend on not only the input signal property, but also the frequency of upstream rhythmic modulator. The frequency modulation of neurons information transferring
 is dependent on the signal strength. Strong signals are more tend to be transferred with high frequency oscillation, whereas weak signals are tend to be
 transferred by slow oscillations. The underlie mechanism are explained with the phase plane analysis of an equivalent deterministic neuron model. We also argued that the information rate for weak signals is greatly enhanced if information are transferred through multiple pathways modulated
by slow oscillations with different phases.

\section*{Model and Method}
\subsection*{LIF Neuron Model}
We use a simple but analytically tractable model of a spiking neuron, the LIF model, with which the action potentials are generated by a threshold process. Suppose $V_{\theta}> V_{rest}$, and when $v(t) <V_{\theta}$, the membrane potential is written as
\begin{eqnarray}
dv(t) &=& -\frac{v(t)-V_{rest}}{\gamma}dt+dI_{osc}(t)+dI_{pulse}(t),
\label{LIFmodel}
\end{eqnarray}
where $v(t)$ is the membrane potential of the neuron, $V_{\theta}$ the threshold, and $V_{rest}$ the resting potential. $\gamma$ is the decay time constant. The neuron fires an action potential when its membrane potential reaches the threshold potential, then the membrane potential stays at the resting potential for a refractory period of $T_{ref}=5~ms$.

We assume the neuron receives synaptic inputs from $N_{s}$ active synapses, each sending Poisson excitatory post-synaptic potential inputs to the neuron with rate
\begin{equation}
\lambda_{E}(t)=\frac{a}{2}(1+cos(2\pi Ft+\phi)),
\end{equation}
where $a$ and $F$ are constant magnitude and temporal frequency. Then the background oscillating synaptic input in Eq.~\ref{LIFmodel} is defined as
\begin{eqnarray}
dI_{osc}(t)=\mu(t)dt+\delta(t)dB_{t},
\end{eqnarray}
where $\mu(t)=\lambda(t)$ and $\sigma^2(t)=\lambda(t)$, $\lambda(t)=\lambda_{E}(t)N_{s}$ is the input rate\cite{Feng2004}.  $B_{t}$ is standard Brownian motion.

Beside the oscillating inputs from the upstream neurons, the LIF neuron also receives a train of pulse inputs  $I_{pulse}(t)$. The pulse duration is $2 ms$ and the inter-pulse intervals are drawn from a Poisson distribution.
\subsection*{Statistics of spike trains }
An usual way to represent the spike train is to use the sequence of intervals between
the spikes, i.e., $(\delta t_1 , \delta t_2 , ..., \delta t_{n} ) = (t_2-t_1, t_3-t_2 , ..., t_{n}-t_{n-1} )$, where $(t_1, t_2 , ...,t_{n})$ is a point process that represents the spiking times of a sequences of spikes. The distribution of inter-spike intervals is then used to measure the  nature of the spike train that a neuron uses, in response to a specific stimulus. The first two moments of the distribution are, the mean inter-spike interval $\mu_{ISI}$ and the variance of the distribution $Var_{ISI}$. Then we defined the coefficient of variation as $CV=\frac{\sqrt {Var_{ISI}}}{\mu_{ISI}}$  to characterize the variability in the inter-spike intervals~\cite{Dayan}.

\subsection*{Measurement of Entropy and Information in Spike Trains }
 We use the `direct method' to measure the entropy of the spike train in
response to a specific stimulus~\cite{PhysRevLett.80.197}. The spike trains are discretized, using bins of
size $\Delta \tau$, into a binary sequence of zeros (no spike) and ones (spike). A sliding window
of size $T$ is applied along the binary sequence to get a sequence of $K$-letter binary `words' ($K = T /\Delta \tau$).
$P (W)$, the probability of the word $W$ to appear in the spike trains is estimated, and the total entropy of the word distribution is calculated as follows,
\begin{equation}
H^{total}_{T}=-\sum_{W}P(W)\log_{2} P(W).
\label{eq_H}
\end{equation}
This
calculation is repeated for different word sizes (different $K's$). Taking the limit of infinitely
long words (and normalizing by the word length), Eq.~\ref{eq_H} gives the total entropy rate,
\begin{equation}
H^{total}=\lim\limits_{T \rightarrow \infty}\frac{1}{T}H^{total}_{T},
\label{Eq_Htotal}
\end{equation}
which measures the information capacity for the spike
train.

The time-dependent word probability distribution at a particular time $t$, $P (W |t)$, is estimated over all the repeated presentations of the stimulus. At each time $t$ we calculate the time-dependent entropy of the words, and then take the average (over all times) of these entropies,
\begin{equation}
H^{noise}_{T}=<-\sum_{W}P(W|t)\log_{2}P(W|t)>_{t},
\label{eq_Hn}
\end{equation}
where $<...>_{t}$ denotes the average over all times $t$. This calculation is repeated for each of the inputs, using different word sizes ($K$ values). Taking the limit of infinitely long words, Eq.~\ref{eq_Hn} gives the spike trains' noise entropy rate,
\begin{equation}
H^{noise}=\lim\limits_{T \rightarrow \infty} H^{noise}_{T},
\end{equation}
which measures how much of the fine structure of the spike trains of the neuron is just noise.

The difference between the neuron's total entropy rate and the noise entropy rate, is the average information rate that the neuron's spike trains encode about the stimulus:
\begin{equation}
I(stimulus, spike~train)=H^{total}-H^{noise}.
\end{equation}
The bin size $\Delta \tau$ was chosen to be $2 ms$ long, which is small enough to keep the fine temporal structure of the spike train within the word sizes used, yet large enough to avoid undersampling problems.

\section*{Results}
\begin{figure*}[t]
\begin{center}
\includegraphics[width=0.95\textwidth]{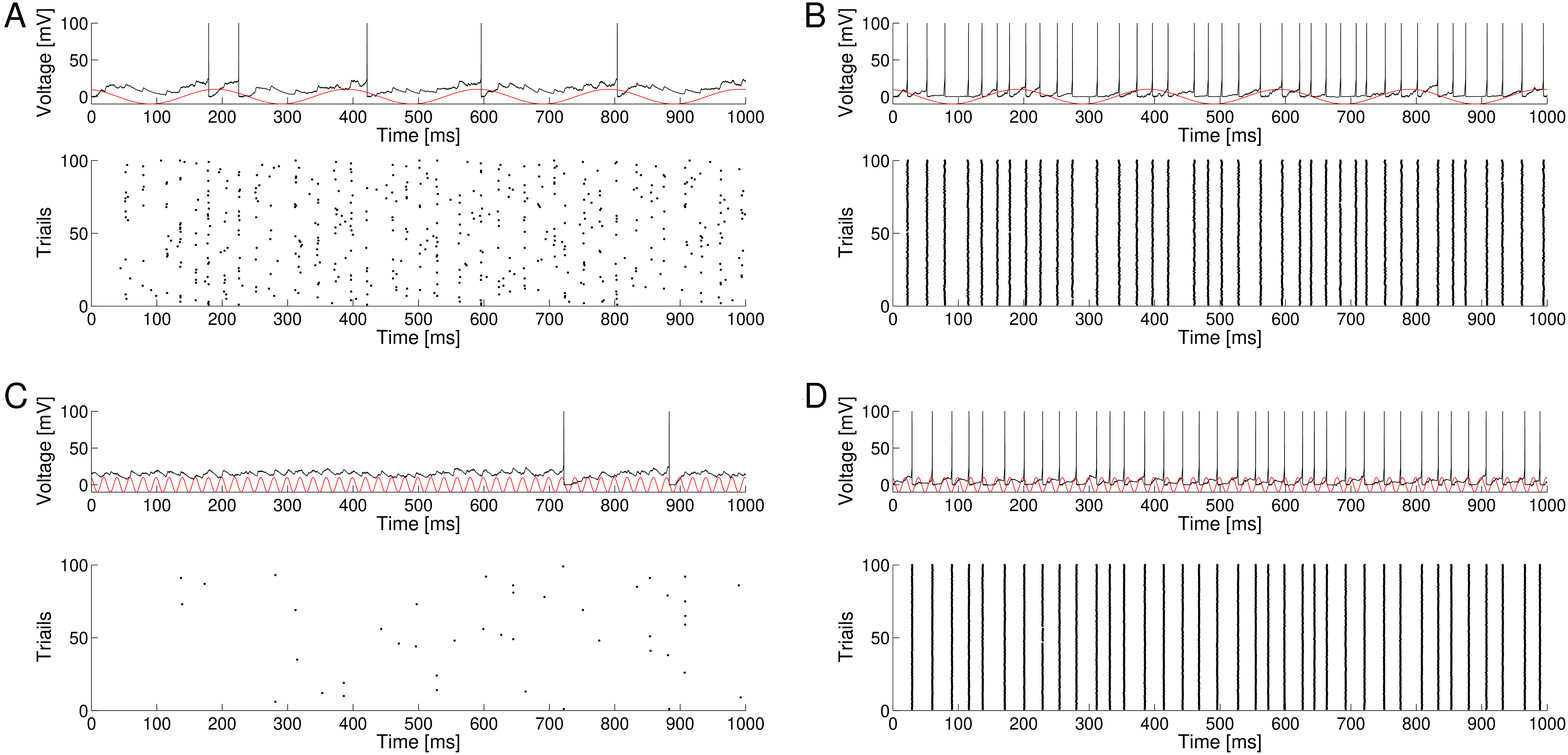}
\end{center}
\caption{ (Color online) Example of the membrane potential traces of a single neuron and spike rasters across $100$ trials in response to
identical input pulse trains for different oscillating frequency of the neuron and pulse strength. The oscillating frequency
of the neuron is $F=5 Hz$ in (A) and (B), and $F=50 Hz$ in (C) and (D). The input pulse strength is $3 \mu A/cm^2$ in (A) and (C),
and $13 \mu A/cm^2$ in (B) and (D). The input pulse rate is $40Hz$ in all plots. Red lines are the corresponding background input waveforms.} \label{fig1}
\end{figure*}

In the following, we consider a LIF neuron whose membrane potential is modulated by $N_{s}$ upstream neurons, each sending Poisson excitatory postsynaptic potential inputs to the neuron with sinusoidal rate. The sinusoidal inputs will cause the oscillation in membrane potential of the LIF neurons, and this oscillating membrane potential will give rise to complex response when another pulse train inputs are applied. The inter-pulse intervals were drawn from a Poisson distribution with rate $\lambda$ and the pulse strength is chosen in the way that the dynamic range of the neuron's response could be covered. Figure.\ref{fig1} demonstrated the spike trains and raster plots the neuron with different oscillating frequency exhibited in response to pulse inputs of different strength.
\subsection*{Spike Train Statistics}

\begin{figure}[t]
\begin{center}
\includegraphics[width=0.5\textwidth]{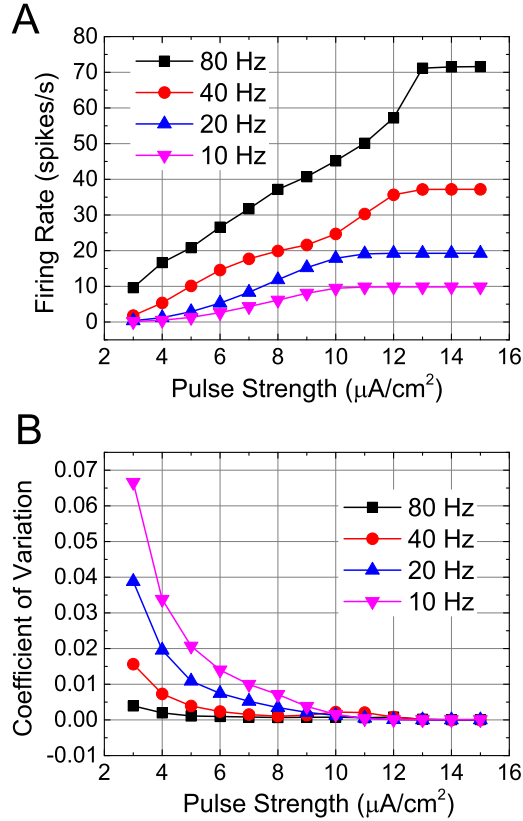}
\end{center}
\caption{ (Color online) Dependence of spike train statistics on the input pulse train parameters. (A) Firing rate as a function of pulse strength for
different pulse rates. (B) Coefficient of variation as a function of pulse strength for different pulse rates. The oscillating frequency of the neuron is
$F=20Hz$.} \label{fig2}
\end{figure}

We first use the mean firing rate and coefficient of variation (CV) to characterize the spike train statistics. Since strong pulses with high input rate induce more spikes in neurons, the firing rate increases as the input rate and strength increases (Fig.~\ref{fig2} A). With the increasing of input strength, the firing rate will increasing and then become constant after it reaches the same value as the input rate (e.g., $10$ and $20$ Hz in Fig.~\ref{fig2} A).
If input rate is high (e.g., $40$ and $80$ Hz in Fig.~\ref{fig2} A), the highest firing rate the neuron could generate is lower than its input rate because the refractory period causes the loss of successive generation of spikes
immediately following a previous firings. Both the increasing of input strength, or input rate could decreasing the coefficient of variation, so to
make the firing events in spike trains more coherent (Fig.~\ref{fig2} B).

\begin{figure}[t]
\begin{center}
\includegraphics[width=0.5\textwidth]{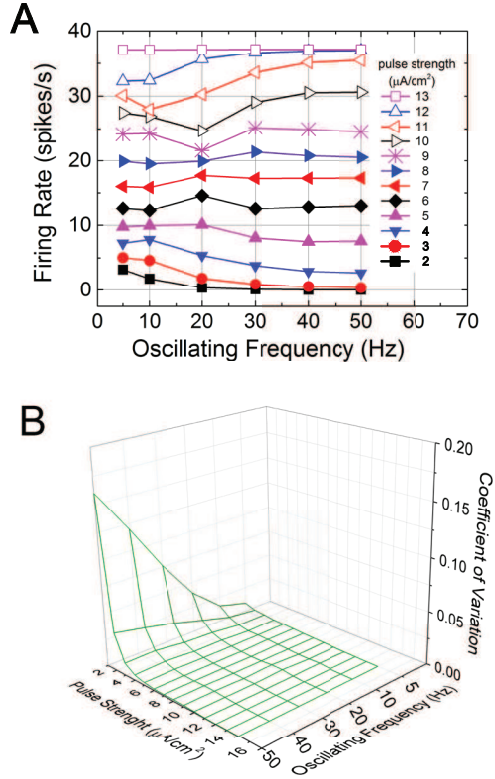}
\end{center}
\caption{ (Color online) The influence of oscillating frequency of the neuron on its spike train statistics in response to pulse trains with different strength.
(A) Firing rate as a function of oscillating frequency for different input strength. (B) Dependence of coefficient of variation on oscillating frequency and
input strength, the data is on the grid and the lines is a guid to the eye.
The input rate is $40Hz$. } \label{fig3}
\end{figure}

As oscillating frequency changes, the firing rate exhibits multiple behaviors, depending on the strength of the inputs the neuron receives (Fig.~\ref{fig3} A).
For extremely strong inputs, the firing rate is independent of oscillating frequency (e.g., $13~\mu A/cm^2$ in Fig.~\ref{fig3} A). As the input strength decreases, the changing of firing rate with increasing oscillating frequency, undergoes a transition from increasing to decreasing. It turns out that for weak signals, firing rate is higher when neuron is oscillating at lower frequency. It is noted that for moderate input strength, the firing rates \emph{vs.} oscillating frequency curves exhibits a sudden
drop(above $8 \mu A/cm^2$) or increase (below $8 \mu A/cm^2$) at $20 Hz$, which is the half of the input rate. We confirmed this resonance phenomenon can happen with different input rates (results not shown). For weak signals, the spike trains generated by neurons with slow oscillation are far from coherent than it is with fast oscillation. As input signals become strong, the difference in CV for different oscillating frequency is
trivial (Fig.~\ref{fig3} B).

\subsection*{Information Rate}

\begin{figure}[t]
\begin{center}
\includegraphics[width=0.8\textwidth]{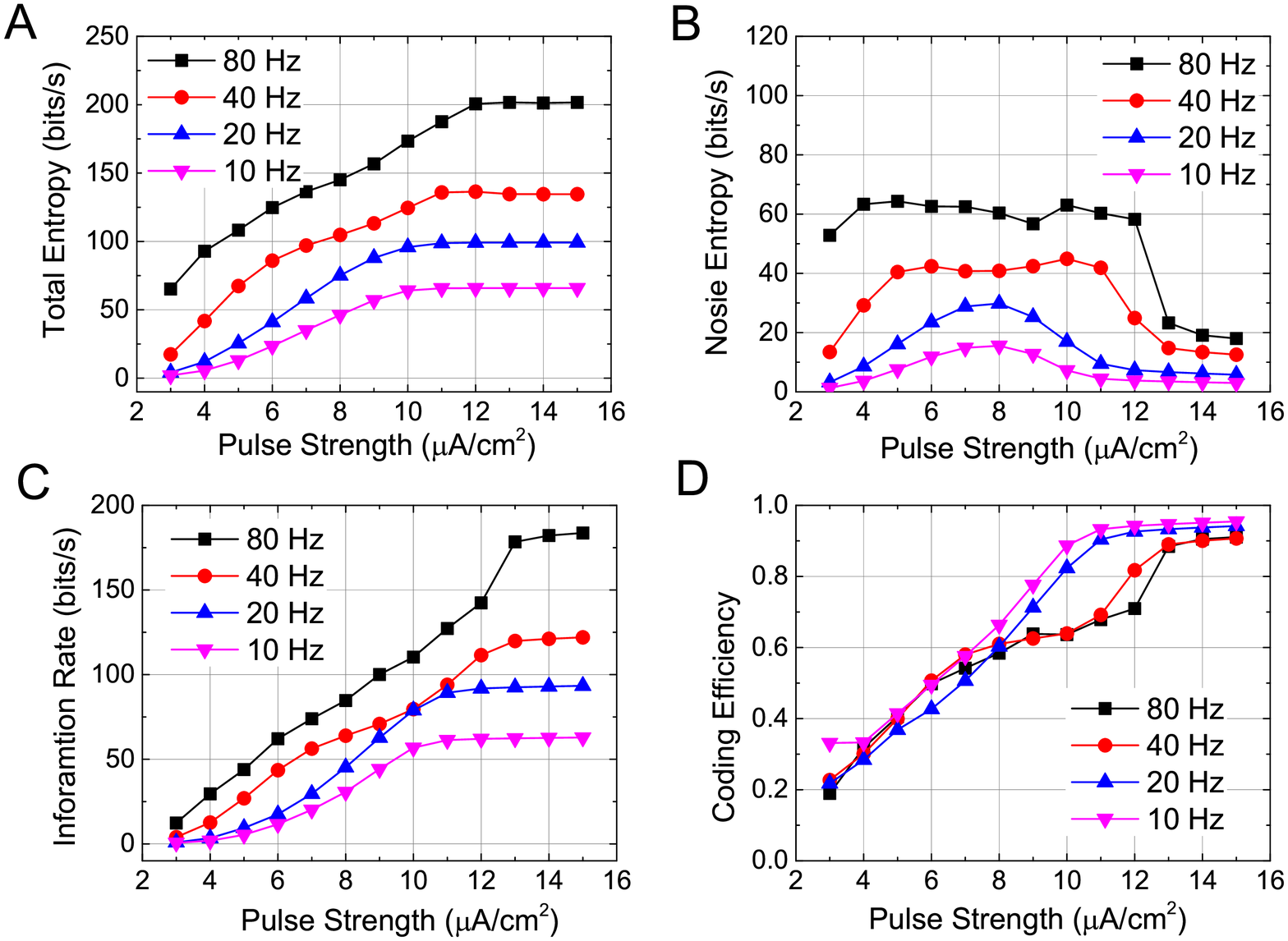}
\end{center}
\caption{ (Color online) The dependence of information capacity for single neuron on the input pulse train parameters. (A)-(D) shows the total entropy, noise entropy, information rate, and coding capacity as a function of pulse strength for different pulse rates, respectively. The oscillating frequency is $F=20Hz$.}
\label{fig4}
\end{figure}

\begin{figure}[6t]
\begin{center}
\includegraphics[width=0.8\textwidth]{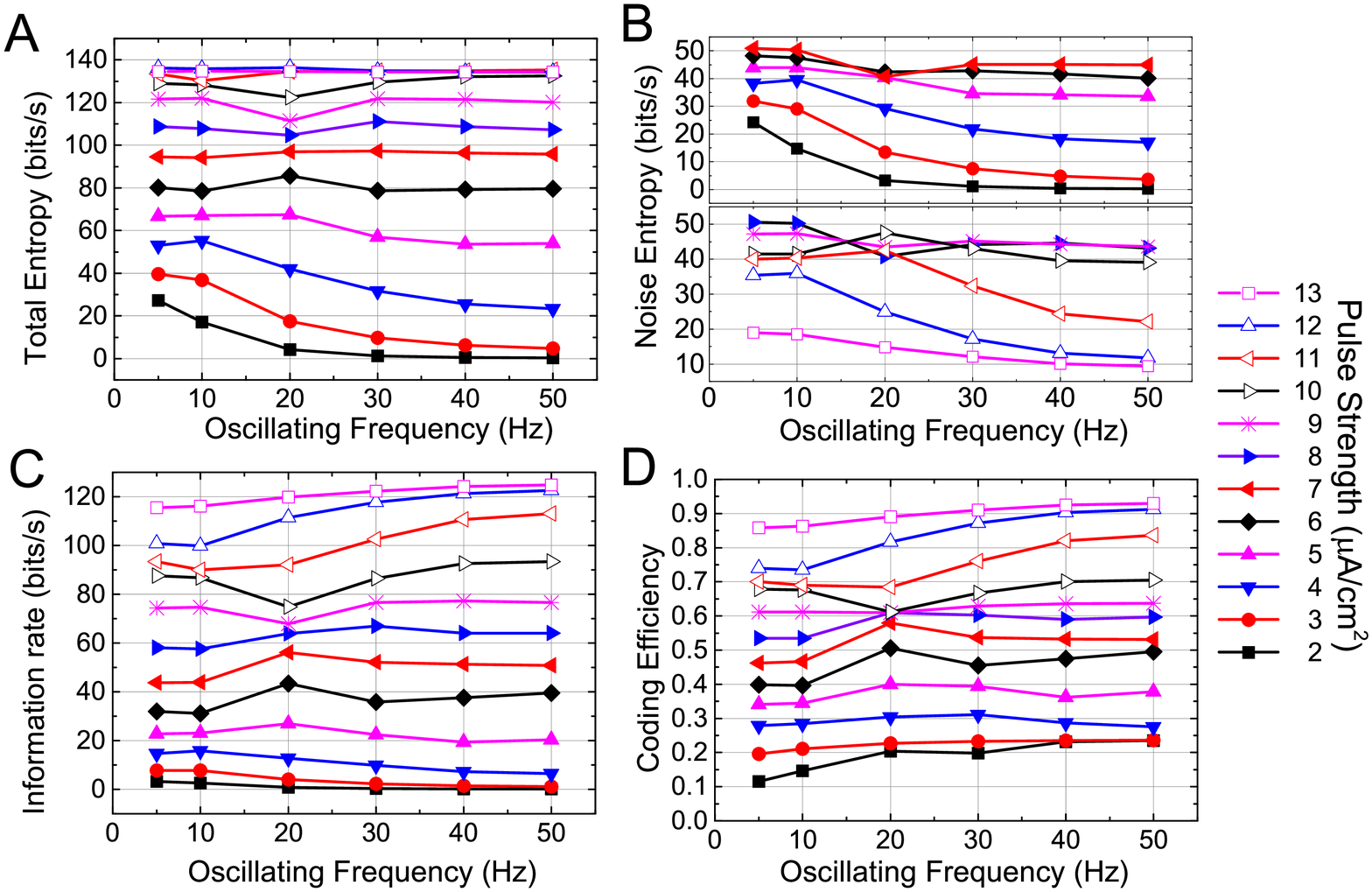}
\end{center}
\caption{ (Color online) The dependence of information capacity for single neuron on the oscillating frequency for different input pulse strength. (A)-(D) shows the total entropy, noise entropy, information rate, and coding capacity as a function of oscillating frequency, respectively. The input pulse rate is $F=40Hz$.}
\label{fig5}
\end{figure}

To quantify the coding properties of the oscillating neuron, we proceed to calculate how much information is conveyed by the spike trains of the model about
each of the stimuli it receives. The total entropy of the spike train gives a bound on how much information it could carry, if there was no noise. Since strong pulses
 with high input rate can make the neuron generate more spikes in unit time, in other words, input more information into the neuron in unit time,
resulting in high total entropy rate of the spike trains generated by neurons in response to them (Fig.~\ref{fig4} A). The dependence of total entropy of
the spike trains on the oscillating frequency was investigated with fixed input rate of $40 Hz$. When pulse strength
is strong ( $6 \mu A/cm^2$ or larger in Fig.~\ref{fig5} A), the oscillating frequency has little effects on the total entropy, except the sudden changes at $20 Hz$.
On the contrary, when pulse strength is weak ( $5 \mu A/cm^2$ or weaker in Fig.~\ref{fig5} A), the total entropy drops as the oscillating frequency increases.

 The noise entropy increases as the input pulse rate increases, and it is high for moderate pulse strength and drops as pulse strength becomes weaker or stronger(Fig.~\ref{fig4} B). As the oscillating frequency increases, the noise entropy decreases substantially for weak or strong inputs, but keeps roughly constant for moderate inputs[Fig.~\ref{fig5} B  (top and down), for better view, the lines corresponding to strong and weak pulses are plotted separately in two graphs]. This phenomena will be explained in the following paragraph.

The information rate, which is the difference between total entropy and noise entropy, increases as pulse strength or input rate increases (Fig.~\ref{fig4} C).
We also noted that with some pulse strengths, increasing in the pulse rate does not bring substantial increasing in information rate. For example, the information
rate changes little when the input rate is increased from $20$ to $40~Hz$ for pulse strength of $10$ or $11 \mu A/cm^2$, because with these pulse strength,
 the noise entropy for input rate of $40~Hz$ is far higher than it is for $20Hz$, and the advantage of high total entropy for $40~Hz$ is offsetted. The changes of information rate on the oscillating frequency of the neurons is also dependent on the pulse strength (Fig.~\ref{fig5} C). For strong pulses, though decreasing the oscillating
frequency will not bring salient changes in the total entropy, the noise entropy decreases. Therefore, the information rate increases as oscillating frequency
increases for strong pulses. For weak pulses, both the total entropy and the noise entropy decreases for increased oscillating frequency, so does the information rate.

We also computed the coding efficiency, which is the ratio between the amount
of information carried by the spike train and the total spike train entropy. The efficiency measure equals one if the the
spike trains are deterministic (noise entropy is zero), in this case all of the spike train structure is utilized to
encode information about the stimulus. The efficiency measure equals zero if the spike train carries no information about the stimulus. Normally, coding efficiency
lies between this two extreme cases. When pulse strength is below $8 \mu A/cm^2$, the coding efficiency is almost the same for different input rate. However,
when pulse strength is larger than $8 \mu A/cm^2$, it turns out the the lower input rate results in higher coding capacity(Fig.~\ref{fig4} D). This phenomenon is
especially significant in the range between $8 \mu A/cm^2$ and $13 \mu A/cm^2$. The coding efficiency increases as oscillating frequency increases, no matter the pulse strength is strong or weak(Fig.~\ref{fig5} D).
The increasing is significant for very strong or very weak pulses, but indistinctively for moderate pulse strength.

\subsection*{Mechanism for Stimulus-Dependent Frequency Modulation}
\begin{figure*}[ht]
\begin{center}
\includegraphics[width=0.8\textwidth]{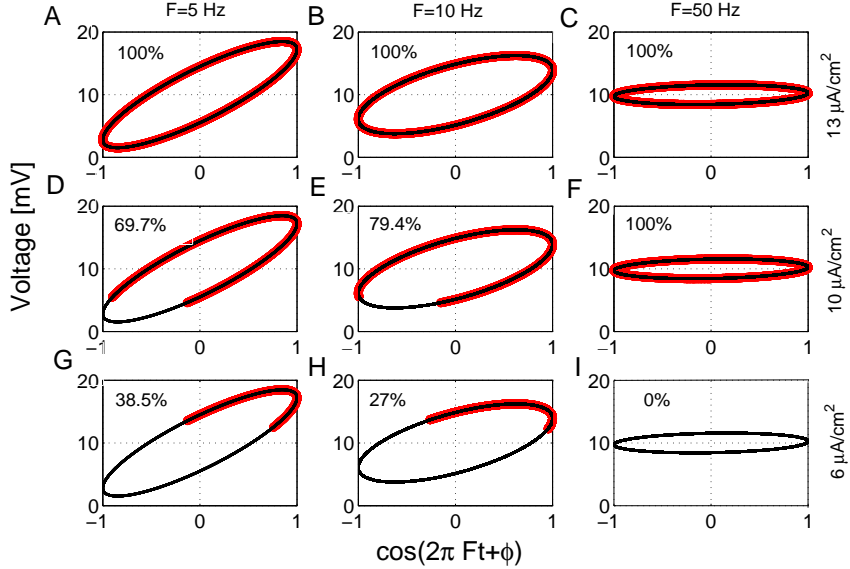}
\end{center}
\caption{ (Color online) Limit cycles of subthreshold oscillation in single neuron with different frequency for the equivalent deterministic model. Red regions mark the `response regions' and percentages indicate the detection rate, as defined in the text.}
\label{fig6}
\end{figure*}

To understand the underline mechanism of above results, let us first discuss a simplified deterministic version of Eq.~\ref{LIFmodel},
\begin{equation}
\frac{dv}{dt} = -(v-V_{rest})/\gamma+C(1+cos(2\pi Ft) ),
\label{deter_lIF}
\end{equation}
where $C=aN_s/2$. Suppose that initially $t=0$, and the membrane potential is reset to $v(0)=V_{rest}$. The theoretical solution for the above equation is
\begin{eqnarray}
v(t) &=& V_{rest}-C \gamma(exp(-\frac{t}{\gamma})-1) \\ \nonumber
&&+ \frac{ C(2\pi F\gamma^2  sin(2\pi Ft)-\gamma exp(-\frac{t}{\gamma})+\gamma cos(2\pi Ft) )}{((2\pi F)^2 \gamma^2+1 )}.
\label{s}
\end{eqnarray}
Thus if time $t$ is infinitely long, the membrane potential $v(t)$ would oscillate around $V_{rest}+C\gamma$ and a limit cycle is seen in the phase space(Fig.~\ref{fig6}), the maximum and minimum values of the membrane voltage are then written as
\begin{equation}
V_{rest}+C\gamma \pm \frac{C\gamma}{\sqrt{((2\pi F)^2 \gamma^2+1)}}.
\end{equation}
Note that as F decreases, the maximum of membrane voltage increases, and the minimum decreased. Thus the limit cycle will be extended along axis of membrane voltage, giving a more tilted limit cycle in the phase space.

When a pulse input is applied, if the system is in the near-threshold part of the limit cycle, it is easy for the system to cross the threshold . On the contrary,
it is difficult or in vain for the system to cross the threshold if it is in the position on the limit cycle that is far away from the threshold.
Since the oscillating frequency influences the position of the limit cycles, the response of the neuron will depend on not only the input strength,
but also the oscillating frequency. To show it more clearly, in Fig.~\ref{fig6} we also plotted with red circles the `response region',
which is corresponding to the collections of the instant states ( points in the phase space). If the system happens to be in one of those states, the threshold-crossing
events happen in less than $5~ms$ after pulse inputs are applied. We can see that if pulse strength is large enough, the response rate is $100\%$(the percentage of firings in response to the inputs in less than $5~ms$, indicated in each plot), independent of oscillating frequency. As pulse strength decreases, neurons
with slow oscillation are easier to lost parts of their `reaction region', because the lower part of the limit cycles is further from the threshold, pulse input
is not able to push the system to cross the threshold. As a result, the response rate increases as oscillating frequency increases. Further decreasing the input strength,
we found the `response region' disappears if oscillation is fast, but upper part of them still keeps if oscillation is slow. As a results, the response rate decreases as the oscillating
frequency increases.

\begin{figure*}[ht]
\begin{center}
\includegraphics[width=0.8\textwidth]{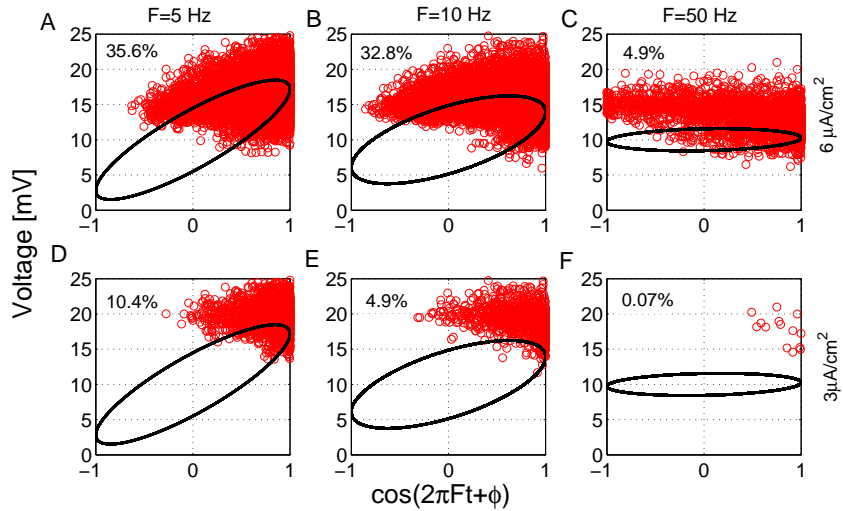}
\end{center}
\caption{ (Color online) Limit cycles of subthreshold oscillation in single neuron with different frequency for the stochastic model. Red regions mark the `response regions' and percentages indicate the corresponding detection rates, as defined in the text.} \label{fig7}
\end{figure*}

This deterministic equivalence of the oscillating neuron will stop to fire if input strength reaches a low boundary. However, with the consideration of its
stochastic part, the neuron will fire again because of the well known phenomenon of `stochastic resonance', which implies that noise could facilitate the
threshold-crossing of weak signals. Again, the limit cycles and the `response regions' are plotted for the system with noise in Fig.~\ref{fig7}.
It is seen that the noise enlarged the `response region' in the upper-part of the limit cycles and enable the neuron to response to weak signals,
and the neurons with slow oscillation are more sensible to the stimulus.

The above analysis explained how the oscillating frequency of the neuron could affect its firing rates and information transmission in response to the pulse inputs. Now let us discuss how the noise entropy is affected by the input strength, as demonstrated in Fig.~\ref{fig4} B. The noise entropy is actually a measurement of the variability in the firing times of neuron. Since the neuron is sure to fire and fire quickly for strong pulses,
the variability of the firing time is low. When the pulse strength is decreased, the subthreshold oscillations will have a great influence on deciding wether or not
the neuron will fire and if the neuron fires, how long the latency to the spike is, giving a high variability of the spiking time. So the noise entropy increases as input strength decreases. Further
decreasing the pulse strength, the neuron can only fire in a narrow `response regions'. This narrow `response regions' would provide narrow distributions of the
 latency time to spikes. We argue this is the reason why the noise entropy drops again for weak signals.

\section*{Discussion}
\begin{figure}[ht]
\begin{center}
\includegraphics[width=0.9\textwidth]{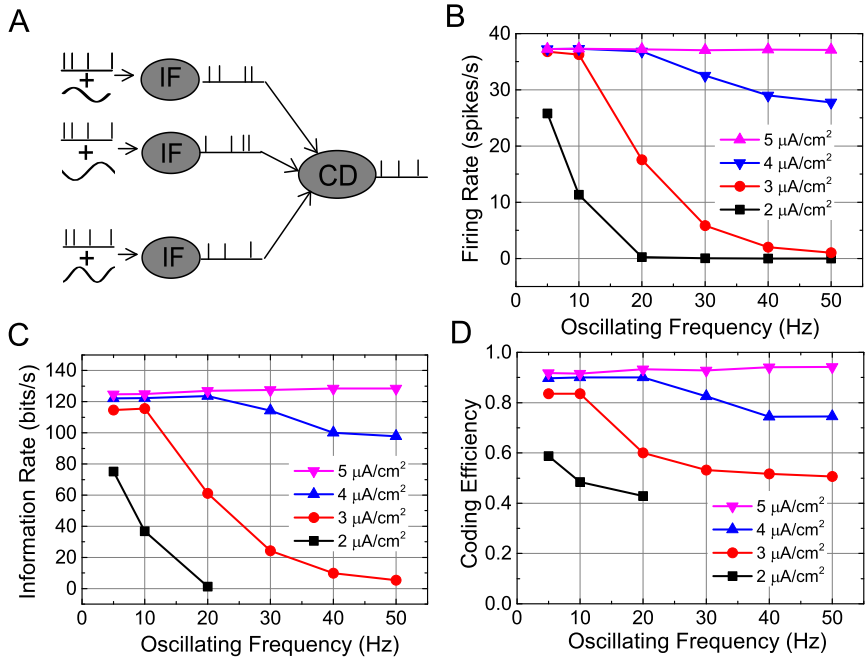}
\end{center}
\caption{ (Color online) Information capacity is enhanced in the multiple neuron pathway. (A) The scheme for multiple pathway neuronal network with CD neuron.
(B)-(D) The firing rate of CD neuron, information rate, and coding capacity as a function of oscillating frequency for different input pulse strength.} \label{fig8}
\end{figure}

In this study, we have investigated how the neuron's subthreshold membrane potential oscillations generated by upstream neuron activities, could modulated
its information processing through oscillating frequency. Our results suggests that information carried by strong signals are more likely to be conveyed by fast oscillations, whereas the slow oscillation facilitate information transmission of weak signals. The underlying mechanism for these results is related to the
 position of the limit cycles corresponding to the subthreshold oscillations. The limit cycles are more tilted for slow oscillation, providing a more excitable
transient state for threshold crossing, resulting in the frequency dependent information transmission capacity of single neuron.

\subsection*{Stimulation Strength Dependent Modulation of Information Transmission with Oscillation Frequency}
 Stimulus dependent
gamma frequency oscillations are often found in primary visual area (V1) of mammalian cortex\cite{Gary1989}. It is reported that the strength of oscillatory synchronization in macaque V1 increases smoothly with stimulus contrast and activities at high contrast show significant gamma frequency modulation~\cite{Henrie2005,Ray2010}. Though there is no strong evidence for low frequency oscillations (\emph{e.g.,} oscillations $<4Hz$ occurs in sleep and anaesthetised state) during stimulation, recent studies in V1 argued that the information is encoded by spike phase relative to low frequency oscillations($<10Hz$), and the information was maximal for the lowest local field potential frequency and decreased monotonically with frequency~\cite{Kayster2009}. Those evidences suggest the possibility that the modulation of information through frequency of oscillatory neural activities,
and the modulation is stimulus dependent.

 Though there are not many studies on the roles of slow oscillation in the neural information processing, several studies show that slow oscillations play an important role in brain function.  The slow dynamics in neural systems can arise from the balance between excitation and inhibition of networks~\cite{Litwin-Kumar}. Studies from resting state functional magnetic resonance imaging (rfMRI) have revealed the important roles of brain slow oscillations (typically between $0.01Hz$ and $0.1 Hz$) in forming of small word nature of whole brain functional connectivity, or the so called default mode networks.  Animal models with reduced synaptic inhibition are characterized by significantly reduced theta oscillation in the hippocampus~\cite{Wulff2009}, and studies suggests that GABA neuron alterations are related to the cortical circuit dysfunction in Schizophrenia~\cite{Guillermo2011}. Recent studies suggests longer inhibition can create slower oscillatory frequency of the neuronal network, and can speed reaction times in a decision-making networks\cite{Smerier2010}. Results in this paper provided a basic understanding of how neural oscillations modulate the information processing in a single neuron level.

\subsection*{Slow Oscillations Facilitate Information Transmission in the Population Level}
One puzzle remained in our results is that in single neuron, though the slow oscillation facilitates the transferring of weak signals, the information rate is very low for weak signals. However, we argue that population coding may be a way out of this predicament. For an example, we constructed a neuronal network as shown in Fig.~\ref{fig8} A.
The neuronal pathway is composed of $100$ LIF oscillating neurons and a downstream coincidence detector (CD) neuron.
The LIF neurons receive the same weak pulse inputs, and background oscillating inputs with same frequency but different phases (the initial phases are distributed evenly in the interval of 0 to 2$\pi$). The outputs of those LIF neurons were taken as the inputs of the CD neuron. The CD neuron fires only when several inputs arrives simultaneously in a short time window. In the simulation, If there are $10$ spikes occurs in a time window of $4~ms$ then the CD
neuron was marked with a firing event at the time the last action potential in the window occurred. After a firing, the CD neuron will enter a refractory period of $10 ms$.

We found that the facilitation of weak signals with slow oscillation is significantly enhanced in this neuronal pathway.
With input strength of $5~\mu A/cm^2$ or above, the firing rate of the CD neuron
reaches the maximal firing rate of a single neuron can have, independent of oscillating frequency (Fig.~\ref{fig8} B).
The information rate and coding efficiency have a slight  increasing as oscillating frequency increases, mainly because the noise entropy is high when oscillating frequency is low (results not shown).
For weaker pulse input, the firing rate, the information rate, and the coding efficiency drop as oscillating frequency increases, and almost reaches the up boundary the single neuron can achieve when oscillating frequency is between $5$ and $10~Hz$ , except for $2 \mu A/cm^2$. For extremely weak pulses, the CD neuron will have little output spikes at high frequency band (\emph{e.g.,} for $2 \mu A/cm^2$, the data is missing
for oscillating frequency larger than $20~Hz$, because the output spikes are very rare so that the entropy values are unable to be calculated).

Coincidence detection is a novel mechanism that the neural systems employed to read out deterministic information from multiple synchronous inputs\cite{10.1371/journal.pone.0056822}. Above example showed that through this CD mechanism, the information processing ability of slow oscillation were greatly enhanced. The only requirement is that the phases of subthreshold oscillation in LIF neurons should be desynchronized. In this case, the upstream oscillating network can gate the information flow in the downstream neuron through modulating its frequency or phases. Indeed, Green and Arduini (1954) reported that hippocampal theta usually occurs together with desynchronized EEG in the neocortex, and hypothesized that the theta is related to arousal, with which the neural system shows greater responsiveness to sensory stimuli~\cite{Green1954}.

\subsection*{Limitation of Our Model and Possible Misconception}
The model we used here is valid under the assumptions that the LIF neuron receives inputs from neurons of upstream modulator network, each sending Poisson EPSPs with a sinusoidal rate~\cite{Tuckwell1988}, and it is capable of explain the diversity of intrinsic frequency encoding patterns in rat cortical neurons~\cite{Kang2010}. However, it did not include detailed neuronal network connections that form the specific neural oscillations observed in experiments~\cite{Kopellbook}. Therefore one should not link our results directly to any specific rhythms, like theta or gamma. Also,  with this model we can not relate the alterations in synapse connections of upstream oscillating neuronal network, and the resulting changes in oscillation frequency, to the information processing capacity of downstream systems, which is a particular interesting topic to the brain diseases studies. Those drawbacks will be overcome in the future with detailed spiking neuronal network model for specific neural oscillations (for an example, see~\cite{xuejiuan2012}).

\section*{Conclusion}

In conclusion, the information processing capacity of a LIF neuron in response to the pulse inputs is studied under the modulation of upstream neural oscillations.  It is found that the modulating depends on the input strength, \emph{i.e.,} fast oscillations are tend to facilitate the processing of strong signals, but slow oscillations the weak signals.
Those results provided a basic understanding of how the brain could modulate its information processing simply through changing its oscillating frequency, instead of changing its structural connections, like synaptic strength, \emph{etc}. Further studies should focus on understanding the modulation of the specific neural oscillation frequency on neural information processing, using the well established neural networks, and the consequence when those networks are malfunctioned in the perspective of brain diseases.

\section*{Acknowledgments}
Conceived and designed the experiments: LCY LFW. Performed the experiments: LCY LFW. Analyzed the data: LCY LFW FJ DJJ. Wrote the paper: LCY LFW FJ DJJ.
\nolinenumbers

%
%
%
\bibliography{ref}

\end{document}